\documentclass[conference,nofonttune]{IEEEtran}
\IEEEoverridecommandlockouts
\usepackage{cite}
\usepackage{amsmath,amssymb,amsfonts}
\usepackage{fontspec}
\usepackage{algorithmic}
\usepackage{algorithm}
\usepackage{graphicx}
\usepackage{textcomp}
\usepackage{xcolor}
\usepackage{booktabs}
\usepackage[letterpaper, top=0.75in, bottom=1in, left=0.625in, right=0.625in]{geometry}

\newtheorem{theorem}{Theorem}

\newtheorem{corollary}{Corollary}
\newtheorem{proposition}{Proposition}
\newtheorem{remark}{Remark}

\def\BibTeX{{\rm B\kern-.05em{\sc i\kern-.025em b}\kern-.08em
    T\kern-.1667em\lower.7ex\hbox{E}\kern-.125emX}}

\begin{document}

\title{Algorithm-Hardware Co-Design of a Lightweight PCG Equalizer with a Fixed Step Size for Massive MIMO}

\author{
\IEEEauthorblockN{Junshuo Wang, Shuting Liu, Shihan Wang, Yan Tian, Jienan Chen\textsuperscript{*}}
\IEEEauthorblockA{\textit{University of Electronic Science and Technology of China, Chengdu, China} \\
\textsuperscript{*}Corresponding author: jesson.chen@outlook.com}
}

\maketitle

\IEEEaftertitletext{\vspace{-2\baselineskip}} 

\begin{abstract}
Coarse quantization in massive multiple-input multiple-output (MIMO) systems reduces power but causes clipping distortions. The Bayesian Expectation-Maximization (BEM) algorithm can recover clipped signals, but its matrix inversion and dynamic step-size evaluation are hardware bottlenecks. We propose a hardware-friendly one-step correction that uses the initial Jacobi-preconditioned Conjugate Gradient (PCG) direction with a fixed relaxation parameter. The resulting symbol-level update has an ultra-lightweight $\mathcal{O}(U)$ feed-forward datapath and approaches high-resolution reference detectors in the evaluated massive-MIMO setting. Our finite-dimensional analysis establishes the exact one-step descent law, proves that Jacobi normalization cancels the raw multiplicative near-far scaling while confining the loaded-system dependence to bounded attenuation factors, and gives verifiable sufficient conditions for fixed-step descent in terms of normalized channel coherence. System-level results indicate projected power savings for energy-efficient massive MIMO uplinks.
\end{abstract}


\vspace{-3mm}
\section{Introduction}
Massive multiple-input multiple-output (MIMO) is the cornerstone of 5G and 6G networks, especially when integrated with orthogonal-frequency-division-multiplexing (OFDM)\cite{bjornson2017massive}. To realize green communications and accommodate the stringent power budgets, a practical approach is to employ low-resolution analog-to-digital converters (ADCs) in base-station\cite{kim2024user}. Unfortunately, aggressive coarse quantization introduces severe non-linear clipping distortions. Unlike independent thermal noise, these quantization artifacts are heavily correlated with the transmitted signal sequences, severely skewing the constellation geometry. Conventional linear detectors, such as the minimum mean-square error (MMSE) equalizer, typically rely on flat Gaussian noise assumptions and can struggle to compensate for these signal-dependent non-linear artifacts \cite{zhao2024massive}.

In contrast, Bayesian Expectation-Maximization (BEM) algorithms approach the theoretical optimum by leveraging discrete constellation priors to reconstruct clipped amplitudes \cite{wen2016bayes}. Despite its superiority, the exact BEM faces a critical implementation bottleneck as it requires a dense matrix inversion with $\mathcal{O}(U^3)$ complexity per symbol, leading to unacceptable silicon area and power dissipation in application-specific integrated circuits (ASICs) \cite{li2025deep}.

To bypass direct inversions, iterative solvers such as Gauss-Seidel (GS) and Conjugate Gradient (CG) are widely adopted \cite{yin2014conjugate,Khoso2024gs}. However, GS detection suffers from strict sequential data dependencies, while for CG, dynamic step-size evaluation typically incurs dense $\mathcal{O}(U^2)$ operations and creates feedback dependencies in pipelined hardware \cite{wu2014large}. Hardcoding the step size enables a feed-forward $\mathcal{O}(U)$ datapath, but can be unstable without justification.

In this paper, we bridge theoretical insight and hardware feasibility via an algorithm--hardware co-design. Our main contributions are: \textit{1) Exact One-Step Analysis:} we derive the optimal relaxation parameter, the exact relative loss caused by a fixed-step mismatch, and the necessary-and-sufficient descent condition for the quadratic recovery surrogate. \textit{2) Deterministic Near-Far Normalization:} we prove that Jacobi scaling cancels the raw $\sqrt{p_ip_j}$ factors in normalized off-diagonal Gram entries; with diagonal loading, the remaining power dependence is confined to attenuation factors in $(0,1]$. This yields checkable spectral and fixed-step guarantees without an asymptotic independence assumption. \textit{3) Hardware-Oriented Reformulation:} we remove the dynamic step-size evaluator, enabling an $\mathcal{O}(U)$ feed-forward symbol-level datapath \cite{liu2020rcgmmse}. \textit{4) System-Level Validation:} in our setup, the 6-bit design achieves 8.558 bps/Hz and indicates a projected net power reduction of 289.1 mW under our evaluation assumptions.

\textit{Notation:} Column vectors and matrices are represented by boldface lowercase and uppercase letters, e.g., $\mathbf{x}$ and $\mathbf{H}$, respectively. $(\cdot)^H$ and $(\cdot)^{-1}$ denote the conjugate transpose and matrix inverse. The symbol $\|\cdot\|_2$ denotes the Euclidean norm for vectors and the spectral norm for matrices. For a vector, $\operatorname{diag}(\cdot)$ forms a diagonal matrix; for a square matrix, it returns the diagonal matrix obtained by zeroing all off-diagonal entries. $\lambda_{\min}(\cdot)$, $\lambda_{\max}(\cdot)$, and $\kappa(\cdot)$ denote the extreme eigenvalues and the spectral condition number of a Hermitian positive-definite matrix. $|\cdot|^2$ applied to a vector denotes element-wise squared magnitude. $\mathbf{e}_i$, $\mathbf{I}$, $\mathbf{0}$, and $\mathbf{1}$ denote the $i$th standard basis vector, the identity matrix, the zero vector or matrix, and the all-ones vector, respectively. $\mathcal{CN}(\mathbf{0},\boldsymbol{\Sigma})$ denotes a circularly symmetric complex Gaussian distribution. Unless otherwise specified, bit-width prefixes (e.g., 6-bit) denote ADC resolution.

 \vspace{-1.5mm}
\section{System Model and Problem Statement}
\vspace{-1.5mm}
\subsection{Quantized Massive MIMO Uplink}
Consider the uplink of a quantized massive MIMO-OFDM system with $B$ receive antennas and $U$ users. On one subcarrier, the antenna-domain quantized observation is
\begin{equation}
    \mathbf{y}_{\rm q}=Q(\mathbf{H}_{\rm phy}\mathbf{x}+\mathbf{n}),
\end{equation}
where $\mathbf{H}_{\rm phy}\in\mathbb{C}^{B\times U}$, $\mathbf{x}\in\mathbb{C}^{U}$, $\mathbf{n}\sim\mathcal{CN}(\mathbf{0},\sigma^2\mathbf{I})$, and $Q(\cdot)$ acts componentwise. Let $\mathbf{F}\in\mathbb{C}^{N_{\rm beam}\times B}$ denote the selected DFT beamformer with orthonormal rows. We define the effective beam-domain observation and channel as $\mathbf{y}\triangleq\mathbf{F}\mathbf{y}_{\rm q}\in\mathbb{C}^{N_{\rm beam}}$ and $\mathbf{H}\triangleq\mathbf{F}\mathbf{H}_{\rm phy}\in\mathbb{C}^{N_{\rm beam}\times U}$. All subsequent equations use these effective beam-domain quantities. The projection reduces Gram-matrix formation to $\mathcal{O}(U^2N_{\rm beam})$ complex multiplications \cite{sayeed2002deconstructing}.

\vspace{-1mm}
\subsection{BEM-Inspired Quadratic Recovery Surrogate}
The exact likelihood of the quantized observation is nonquadratic. The recovery stage considered here instead uses a decision-directed quadratic surrogate motivated by BEM updates. Conditioned on the current hard-decision vector $\bar{\mathbf{x}}$ and its diagonal weight matrix $\mathbf{W}$, the surrogate is
\begin{equation}
    J(\mathbf{x}) = \|\mathbf{y} - \mathbf{H}\mathbf{x}\|_2^2 + (\mathbf{x} - \bar{\mathbf{x}})^H \mathbf{W} (\mathbf{x} - \bar{\mathbf{x}}) + \sigma^2 \|\mathbf{x}\|_2^2,
\end{equation}
where the penalty weight matrix is
\begin{equation}
    \mathbf{W} = \operatorname{diag}(|\bar{\mathbf{x}}|^2 + \delta_{\rm w}\mathbf{1}),
\end{equation}
where $\delta_{\rm w}>0$ is the weight floor that ensures strictly positive diagonal loading. Minimizing $J(\mathbf{x})$ requires solving $\mathbf{A}\mathbf{x}=\mathbf{b}$ with
\begin{equation}
\mathbf{A}=\mathbf{H}^H\mathbf{H}+\mathbf{D},\qquad
\mathbf{b}=\mathbf{H}^H\mathbf{y}+\mathbf{W}\bar{\mathbf{x}},\qquad
\mathbf{D}=\mathbf{W}+\sigma^2\mathbf{I}.
\end{equation}
Here $\mathbf{G}\triangleq\mathbf{H}^H\mathbf{H}$ and $\mathbf{A},\mathbf{G},\mathbf{D},\mathbf{W}\in\mathbb{C}^{U\times U}$, while $\mathbf{b}\in\mathbb{C}^{U}$.
All formal descent statements below concern this conditioned quadratic surrogate; they are not claims about the exact quantized-data likelihood, spectral efficiency, or multi-iteration convergence.

\textit{Problem Statement:} The block-level computation of the Gram matrix $\mathbf{G}=\mathbf{H}^H\mathbf{H}$ and its $\mathcal{O}(U^3)$ factorization (used for ZF initialization) is amortized over the coherence block. In contrast, the diagonal loading $\mathbf{D}$ is symbol-level fast-varying, so repeatedly inverting or refactorizing $\mathbf{A}=\mathbf{G}+\mathbf{D}$ per symbol becomes the real-time bottleneck. Therefore, our goal is not to remove all $\mathcal{O}(U^3)$ costs, but to avoid repeated inversion of $\mathbf{A}$ and to eliminate the per-symbol $\mathcal{O}(U^2)$ dynamic step-size evaluation in CG, enabling an $\mathcal{O}(U)$ feed-forward symbol-level update with a fixed step size.

\begin{algorithm}[!t]
\caption{One-Step Fixed-Step Jacobi-Preconditioned Correction}
\label{alg:BEM_PCG_Simplified}
\small
\renewcommand{\baselinestretch}{1.2}\selectfont 
\begin{algorithmic}[1]
    \REQUIRE $\mathbf{H}, \mathbf{y}, \sigma^2, \delta_{\rm w}, \alpha_{\rm fix}$
    \ENSURE $\hat{\mathbf{x}}$
    
    \STATE $\mathbf{G} = \mathbf{H}^H\mathbf{H}$ and $\mathbf{z} = \mathbf{H}^H\mathbf{y}$
    \STATE Obtain initial ZF solution: $\hat{\mathbf{x}}^{(0)} = \mathbf{G}^{-1}\mathbf{z}$
    \STATE Perform hard-decision: $\bar{\mathbf{x}} = \operatorname{dec}(\hat{\mathbf{x}}^{(0)})$
    \STATE Define $\mathbf{W} = \operatorname{diag}(|\bar{\mathbf{x}}|^2 + \delta_{\rm w}\mathbf{1})$ and $\mathbf{D} = \mathbf{W} + \sigma^2 \mathbf{I}$
    \STATE Initial residual: $\mathbf{r}^{(0)} = \mathbf{W}\bar{\mathbf{x}} - \mathbf{D}\hat{\mathbf{x}}^{(0)}$
    \STATE Jacobi Preconditioning: $\mathbf{p}^{(0)} =[\operatorname{diag}(\mathbf{G} + \mathbf{D})]^{-1} \mathbf{r}^{(0)}$
    \STATE Fixed-Step-Size Signal update: $\hat{\mathbf{x}} = \hat{\mathbf{x}}^{(0)} + \alpha_{\rm fix} \mathbf{p}^{(0)}$
    
    \RETURN $\hat{\mathbf{x}}$
\end{algorithmic}
\end{algorithm}

The direction in line 6 is exactly the initial PCG search direction. Because only one update is executed and the exact PCG line search is replaced by a fixed relaxation parameter, no conjugacy across iterations is invoked.

\section{Formal Analysis of the One-Step Correction}
The analysis is conditioned on the effective channel $\mathbf{H}$, observation $\mathbf{y}$, hard decision $\bar{\mathbf{x}}$, and weight matrix $\mathbf{W}$. It is therefore finite-dimensional and does not require the quantization error or the residual to be Gaussian or independent of the channel.

\subsection{Well-Posedness, Residual Cancellation, and Exact Line Search}

\begin{proposition}[HPD surrogate and exact residual cancellation]
\label{prop:spd_residual}
Let $\delta_{\rm w}>0$ and assume that $\mathbf{H}$ has full column rank, so that $\mathbf{G}=\mathbf{H}^H\mathbf{H}\succ0$. Then $\mathbf{A}=\mathbf{G}+\mathbf{D}\succ0$, the surrogate $J$ has the unique minimizer $\mathbf{A}^{-1}\mathbf{b}$, and the ZF initializer $\hat{\mathbf{x}}^{(0)}=\mathbf{G}^{-1}\mathbf{H}^H\mathbf{y}$ satisfies
\begin{equation}
\mathbf{r}^{(0)}\triangleq\mathbf{b}-\mathbf{A}\hat{\mathbf{x}}^{(0)}
=\mathbf{W}\bar{\mathbf{x}}-\mathbf{D}\hat{\mathbf{x}}^{(0)}.
\label{eq:residual_cancellation}
\end{equation}
\end{proposition}

\begin{remark}[Finite-precision initializer]
\label{rem:finite_precision}
The cancellation in \eqref{eq:residual_cancellation} assumes an exact ZF solve. If $\mathbf{e}_0\triangleq\mathbf{H}^H\mathbf{y}-\mathbf{G}\hat{\mathbf{x}}^{(0)}\ne\mathbf{0}$, then the true residual is
\begin{equation}
\mathbf{r}_{\rm true}^{(0)}=\mathbf{e}_0+\mathbf{W}\bar{\mathbf{x}}-\mathbf{D}\hat{\mathbf{x}}^{(0)}.
\label{eq:finite_precision_residual}
\end{equation}
The descent results below apply directly to $\mathbf{r}_{\rm true}^{(0)}$. If hardware instead uses the simplified residual $\mathbf{r}_{\rm impl}^{(0)}=\mathbf{r}_{\rm true}^{(0)}-\mathbf{e}_0$, then, with $\mathbf{M}=\operatorname{diag}(\mathbf{A})$, its direction is descending only when $\operatorname{Re}\{(\mathbf{r}_{\rm true}^{(0)})^H\mathbf{M}^{-1}\mathbf{r}_{\rm impl}^{(0)}\}>0$; this condition is not established by the present synthesis data.
\end{remark}

\begin{proposition}[One-step line search and mismatch loss]
\label{prop:line_search}
Let $\mathbf{M}=\operatorname{diag}(\mathbf{A})$, $\mathbf{r}=\mathbf{r}^{(0)}\ne\mathbf{0}$, and $\mathbf{p}=\mathbf{M}^{-1}\mathbf{r}$. For real $\alpha$, define $f(\alpha)=J(\hat{\mathbf{x}}^{(0)}+\alpha\mathbf{p})$. Its unique minimizer is
\begin{equation}
\alpha^*(\mathbf{r})=
\frac{\mathbf{r}^H\mathbf{M}^{-1}\mathbf{r}}
{\mathbf{r}^H\mathbf{M}^{-1}\mathbf{A}\mathbf{M}^{-1}\mathbf{r}}>0.
\label{eq:exact_alpha}
\end{equation}
Moreover,
\begin{equation}
\eta(\alpha)\triangleq
\frac{f(\alpha)-f(\alpha^*)}{f(0)-f(\alpha^*)}
=\left(\frac{\alpha-\alpha^*}{\alpha^*}\right)^2,
\label{eq:sensitivity_law}
\end{equation}
and $\alpha$ gives strict one-step descent if and only if
\begin{equation}
0<\alpha<2\alpha^*(\mathbf{r}).
\label{eq:residual_descent}
\end{equation}
\end{proposition}

The proofs of Propositions \ref{prop:spd_residual} and \ref{prop:line_search} are given in Appendix \ref{app:exact_descent}. Equation \eqref{eq:sensitivity_law} measures lost descent of the quadratic surrogate only; it is not an error-rate or spectral-efficiency bound.

\subsection{Why an Unnormalized Global Step Is Fragile}

\begin{proposition}[Near-far counterexample without diagonal normalization]
\label{prop:unscaled_fragility}
Consider the diagonal family $\mathbf{A}=\mathbf{P}=\operatorname{diag}(p_1,\ldots,p_U)$ with $0<p_{\min}\le p_i\le p_{\max}$, and use the unpreconditioned direction $\mathbf{p}=\mathbf{r}$. For a residual aligned with user $i$, the optimal step is $\alpha_i^*=1/p_i$. Consequently,
\begin{equation}
\min_{\alpha>0}\max_{p\in[p_{\min},p_{\max}]}
(\alpha p-1)^2
=\left(\frac{\kappa_P-1}{\kappa_P+1}\right)^2,
\quad \kappa_P\triangleq\frac{p_{\max}}{p_{\min}},
\label{eq:unscaled_minimax}
\end{equation}
with minimizer $\alpha=2/(p_{\min}+p_{\max})$. The worst-case relative loss tends to one as $\kappa_P\to\infty$. In addition, any $\alpha\ge2/p_{\max}$ is non-descending for the strongest-user direction.
\end{proposition}

Proposition \ref{prop:unscaled_fragility} is a counterexample to a uniform unpreconditioned fixed-step guarantee; it does not assert that every ill-conditioned realization diverges. Its proof is in Appendix \ref{app:unscaled_fragility}.

\subsection{Exact Removal of Near-Far Power Scaling by Jacobi Normalization}

Factor the effective channel as $\mathbf{H}=\mathbf{Z}\mathbf{P}^{1/2}$, where $\mathbf{P}=\operatorname{diag}(p_1,\ldots,p_U)\succ0$ contains the user powers and the nonzero columns of $\mathbf{Z}\in\mathbb{C}^{N_{\rm beam}\times U}$ contain all remaining small-scale fading and correlation. Define
\begin{equation}
\mathbf{C}=\mathbf{Z}^H\mathbf{Z},\quad
\boldsymbol{\Delta}=\operatorname{diag}(\mathbf{C}),\quad
\mathbf{T}=\boldsymbol{\Delta}^{-1/2}\mathbf{C}\boldsymbol{\Delta}^{-1/2}.
\label{eq:normalized_gram}
\end{equation}
The matrix $\mathbf{T}$ is the normalized Gram matrix of the user-channel directions and has unit diagonal.

\begin{theorem}[Exact Jacobi normalization identity]
\label{thm:jacobi_identity}
Let $\mathbf{D}=\operatorname{diag}(d_1,\ldots,d_U)\succeq0$, $\mathbf{A}=\mathbf{P}^{1/2}\mathbf{C}\mathbf{P}^{1/2}+\mathbf{D}$, $\mathbf{M}=\operatorname{diag}(\mathbf{A})$, and $\widetilde{\mathbf{A}}=\mathbf{M}^{-1/2}\mathbf{A}\mathbf{M}^{-1/2}$. Define
\begin{equation}
\boldsymbol{\Gamma}=\operatorname{diag}(\gamma_1,\ldots,\gamma_U),
\qquad
\gamma_i=\frac{p_i C_{ii}}{p_i C_{ii}+d_i}\in(0,1].
\end{equation}
Then the following finite-dimensional identity holds exactly:
\begin{equation}
\widetilde{\mathbf{A}}
=\mathbf{I}+\boldsymbol{\Gamma}^{1/2}(\mathbf{T}-\mathbf{I})
\boldsymbol{\Gamma}^{1/2}.
\label{eq:jacobi_identity}
\end{equation}
Hence
\begin{equation}
\|\widetilde{\mathbf{A}}-\mathbf{I}\|_2
\le \|\mathbf{T}-\mathbf{I}\|_2.
\label{eq:jacobi_contraction}
\end{equation}
If $\mathbf{D}=\mathbf{0}$, then $\widetilde{\mathbf{A}}=\mathbf{T}$ and the user-power matrix $\mathbf{P}$ cancels exactly. For $\mathbf{D}\succ0$, the factors $0<\gamma_i<1$ can only contract the deviation from identity.
\end{theorem}

Theorem \ref{thm:jacobi_identity} is proved in Appendix \ref{app:jacobi_identity}. Unlike the previous asymptotic argument, it remains valid for finite $N_{\rm beam}$ and $U$, arbitrary near-far powers, arbitrary channel correlation contained in $\mathbf{Z}$, and the actual positive diagonal loading.

\subsection{Spectral and Fixed-Step Guarantees}

Let $\varepsilon\triangleq\|\mathbf{T}-\mathbf{I}\|_2$. Combining Theorem \ref{thm:jacobi_identity} with the Rayleigh quotient in \eqref{eq:exact_alpha} gives the following result.

\begin{corollary}[Verifiable fixed-step guarantee]
\label{cor:fixed_step}
If $\varepsilon<1$, then, for every nonzero residual,
\begin{equation}
\frac{1}{1+\varepsilon}\le\alpha^*(\mathbf{r})
\le\frac{1}{1-\varepsilon}.
\label{eq:alpha_spectral_interval}
\end{equation}
For any fixed $\alpha_{\rm fix}>0$,
\begin{equation}
\eta(\alpha_{\rm fix})\le
\max\!\left\{
[\alpha_{\rm fix}(1-\varepsilon)-1]^2,
[\alpha_{\rm fix}(1+\varepsilon)-1]^2
\right\}.
\label{eq:fixed_step_bound}
\end{equation}
The unit step $\alpha_{\rm fix}=1$ minimizes this worst-case bound and satisfies $\eta\le\varepsilon^2$. It also gives strict descent for every nonzero residual. More generally, a fixed step is uniformly descending whenever
\begin{equation}
0<\alpha_{\rm fix}<\frac{2}{1+\varepsilon}.
\label{eq:uniform_descent}
\end{equation}
\end{corollary}

\begin{corollary}[Favorable-propagation limit]
\label{cor:favorable_limit}
For any sequence of systems satisfying $\|\mathbf{T}-\mathbf{I}\|_2\to0$, one has $\|\widetilde{\mathbf{A}}-\mathbf{I}\|_2\to0$ and $\alpha^*(\mathbf{r})\to1$ uniformly over all nonzero residual directions. This conclusion is independent of the near-far powers $p_i$.
\end{corollary}

The condition in Corollary \ref{cor:favorable_limit} is a precise spectral form of favorable propagation. It is deliberately stated as a verifiable condition rather than inferred from the invalid fixed-load approximation $\mathbf{H}^H\mathbf{H}/N_{\rm beam}\approx\mathbf{I}$.

\begin{proposition}[Minimax calibration over a trusted interval]
\label{prop:minimax_interval}
If only the interval uncertainty $\alpha^*\in[\ell,u]$ with $0<\ell\le u$ is trusted, then the fixed step that minimizes the worst-case envelope of \eqref{eq:sensitivity_law} over every value in that interval is
\begin{equation}
\alpha_{\rm mm}=\frac{2\ell u}{\ell+u},\qquad
\eta_{\rm mm}=\left(\frac{u-\ell}{u+\ell}\right)^2.
\label{eq:minimax_interval}
\end{equation}
\end{proposition}
For the rounded empirical 95\% interval $[0.971,1.187]$ in Table \ref{tab:step_size}, \eqref{eq:minimax_interval} gives the interval-robust calibration $\alpha_{\rm mm}\approx1.068$. The implemented value $\alpha_{\rm fix}=1.075$ is therefore described as an empirical calibration, not as a theoretically unique or population-optimal constant. The interval and its associated loss are finite-sample observations, not deterministic guarantees outside the evaluated population.

\section{Proposed Hardware Architecture}
Motivated by the exact Jacobi normalization identity and the measured step-size distribution in Section III, we propose a hardware-oriented architecture. The design does not reduce the block-level cost of Gram-matrix formation; instead, it targets the fast-varying symbol-level solver path by eliminating the dynamic step-size evaluator.
\vspace{-2mm}
\subsection{Algorithmic Simplifications for Hardware}
Before mapping to the hardware datapath, the detector is algebraically reformulated to eliminate redundant operations:
\begin{itemize}
    \setlength{\itemsep}{2.5pt}
    \item \textit{Cancellation of Dense Matrix-Vector Multiplication:} Since the initial solution is exactly the ZF output, the standard initial residual $\mathbf{r}^{(0)} = \mathbf{b} - \mathbf{A}\hat{\mathbf{x}}^{(0)}$ simplifies to $\mathbf{r}^{(0)} = \mathbf{W}\bar{\mathbf{x}} - \mathbf{D}\hat{\mathbf{x}}^{(0)}$ where $\mathbf{W,D}$ are both diagonal. This algebraic cancellation completely eliminates the $\mathcal{O}(U^2)$ operations required for $\mathbf{G}\hat{\mathbf{x}}^{(0)}$, collapsing the initialization into a lightweight $\mathcal{O}(U)$ element-wise operation.
    \item \textit{Implicit Symmetric Preconditioning:} By executing $\mathbf{p}^{(0)} = \mathbf{M}^{-1} \mathbf{r}^{(0)}$ instead of the symmetric preconditioned direction $\tilde{\mathbf{p}}^{(0)} = \mathbf{M}^{-1/2} \mathbf{r}^{(0)}$, the algorithm implicitly solves the SPD system while circumventing the hardware-prohibitive square-root matrix $\mathbf{M}^{-1/2}$.
\end{itemize}

\vspace{-1.5mm}

\begin{figure*}[t!]
\centerline{\includegraphics[width=0.8\textwidth]{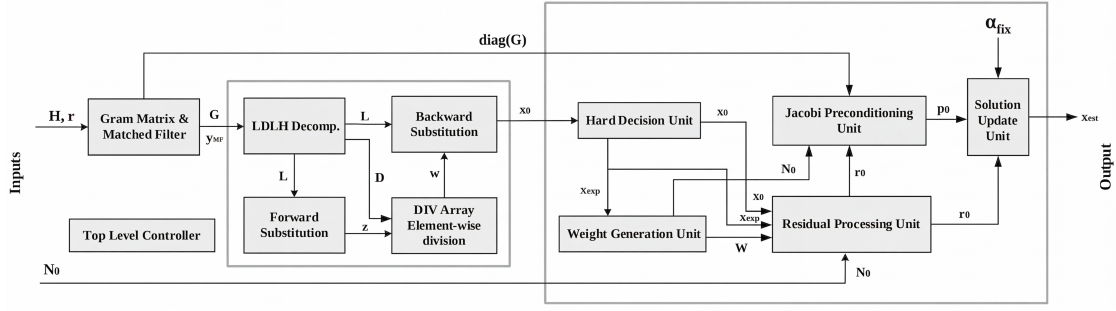}}
\vspace{-3mm}
\caption{Top-level hardware architecture of the proposed one-step fixed-step-size BEM-PCG equalizer.}
\label{fig:arch}
\vspace{-3mm}
\end{figure*}

\subsection{Pipelined Macro-Architecture}
As illustrated in Fig. \ref{fig:arch}, the proposed BEM-PCG equalizer adopts a deeply pipelined macro-architecture orchestrated by a Top-Level Controller (TLC). The architecture decouples slow-varying channel-matrix computations from fast-varying symbol-level updates to improve hardware utilization.

\begin{itemize}
    \item \textit{Precomputation Unit (PCU):} Computes the Gram matrix $\mathbf{G}$ and the matched filter output $\mathbf{y}_{\text{MF}}$. It utilizes a parameterized 2D systolic array with a ``time-for-space'' trade-off.
    \item \textit{Matrix Inversion Unit (MIU):} Solves the initial ZF solution $\hat{\mathbf{x}}^{(0)}$ via a square-root-free $\mathbf{L}\mathbf{D}\mathbf{L}^H$ decomposition. Operating in a slow clock domain under the block-fading assumption, its $\mathcal{O}(U^3)$ complexity is amortized over the entire coherence block.
    \item \textit{Vector/Scalar Processing Unit (VSPU):} Executes the symbol-level one-step PCG updates. It comprises a weight generation unit, a residue processing unit and a hard-decision unit based on arithmetic rounding which is look-up-table (LUT) free. All internal modules rely exclusively on $\mathcal{O}(U)$ element-wise operations.
\end{itemize}

\textit{Hardware Elimination of the Step-Size Evaluator:} Standard CG hardware requires an $\mathcal{O}(U^2)$ matrix-vector-multiplication operation and a high-latency scalar divider for step-size evaluation. For the one-step correction, the exact descent analysis and the calibrated interval in Section III permit this unit to be removed from our VSPU in the evaluated operating regime. As shown in Fig. \ref{fig:arch}, the solution update employs a simple scalar-vector multiplier with $\alpha_{\rm fix}$, enabling a feed-forward $\mathcal{O}(U)$ datapath.

\section{Simulation Results}

We simulate a quantized massive MIMO-OFDM uplink with $B = 2048$, $N_{\text{beam}} = 256$, and $U = 32$. The channel follows the 3GPP CDL-B model with spatial correlation $\rho = 0.2$ and 20 dB near-far effect. Practical channel estimation is implemented utilizing comb-type Zadoff-Chu pilots, Least Squares (LS) estimation, and PCHIP interpolation over the 6-bit quantized beam-domain signals. To emulate realistic 5G network traffic where users experience diverse path losses and employ Adaptive Modulation and Coding (AMC) \cite{3gpp2017ts38211}, a mixed-QAM scheduling scheme (15\% 64-QAM, 40\% 256-QAM, 45\% 1024-QAM) is adopted. Based on this, we compare the proposed 6-bit Fixed-Step-Size BEM-PCG with three 6-bit algorithmic references, namely (i) ideal dynamic-step BEM-PCG, (ii) ideal dynamic-step BEM-CG, and (iii) ideal BEM-GS, alongside two 12-bit high-resolution reference detectors, namely (iv) MMSE and (v) non-linear PIC-MMSE.

\begin{figure}[!t]
\centering
\includegraphics[width=0.8\columnwidth]{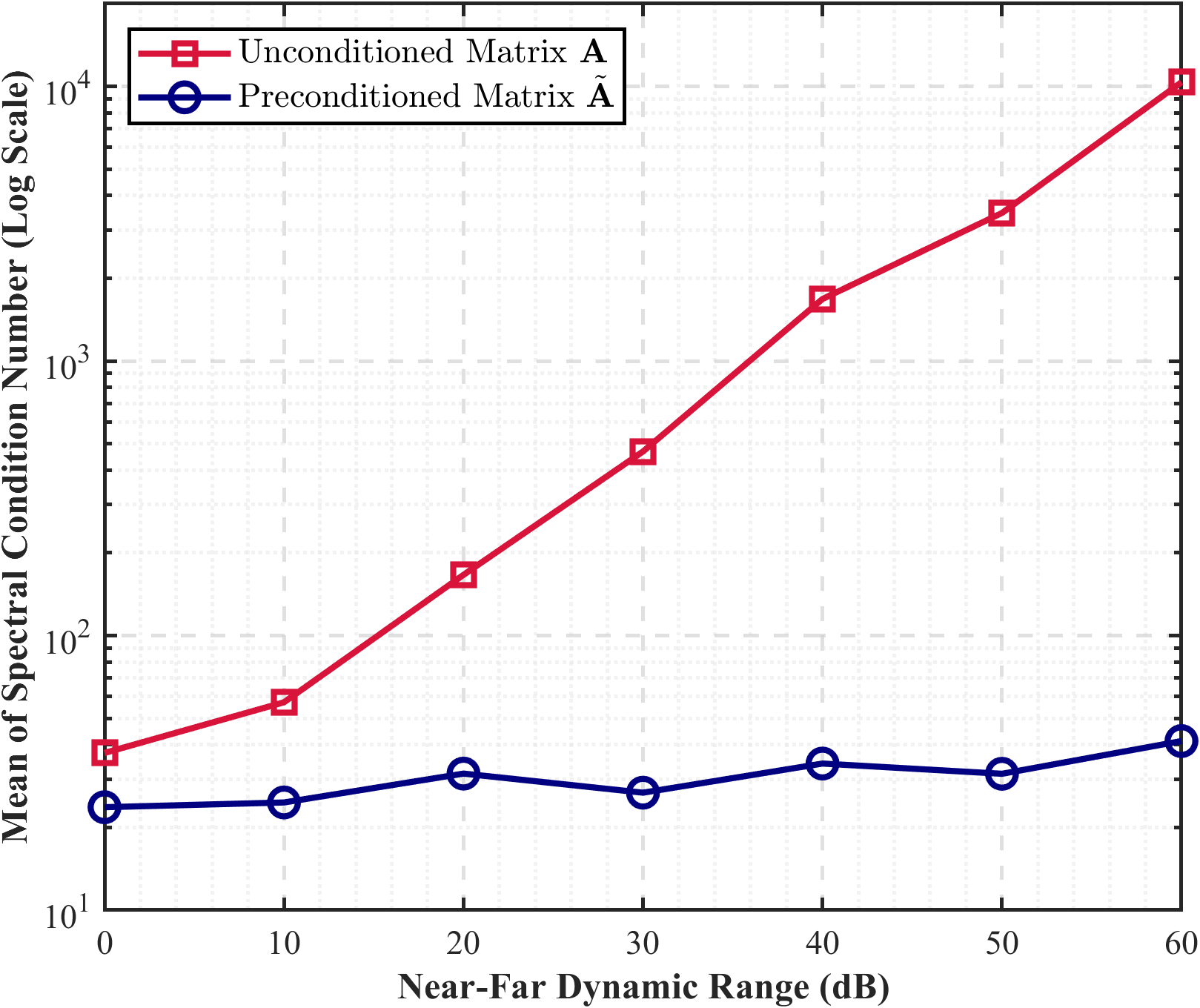}
\vspace{-2mm} 
\caption{Log-scale spectral condition number versus near-far dynamic range for the unconditioned matrix $\mathbf{A}$ and the preconditioned matrix $\widetilde{\mathbf{A}}$.}
\label{fig:cond_num}
\vspace{-2mm}
\end{figure}

\begin{figure}[!t]
\centering
\includegraphics[width=0.8\columnwidth]{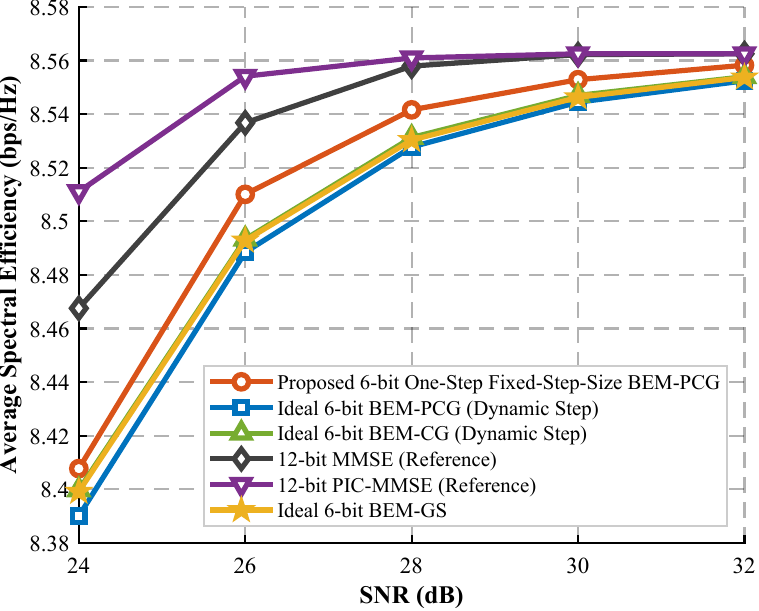}
\caption{Average spectral efficiency (SE) versus SNR for different MIMO detection schemes. $B=2048$; $N_{\text{beam}}=256$; $U=32$; mixed-QAM.}
\label{fig:se_curves}
\end{figure}

\subsection{Condition Number Dynamics and Consistency with the Formal Analysis}
Fig. \ref{fig:cond_num} reports finite-sample condition-number statistics. The unconditioned $\kappa(\mathbf{A})$ grows strongly with the near-far dynamic range, whereas Jacobi normalization keeps the observed $\kappa(\widetilde{\mathbf{A}})$ between 23.7 and 41.3 over the evaluated range. This is consistent with Theorem \ref{thm:jacobi_identity}: the explicit factors $p_i$ cancel from the normalized Gram couplings, while the remaining spread is governed by the normalized channel geometry $\mathbf{T}$ and the loading factors $\gamma_i$. The theorem does not predict the numerical height of the observed plateau.

\begin{table}[!t]
\caption{Empirical Step-Size Statistics ($U=32, \rho=0.2$)}
\label{tab:step_size}
\centering
\renewcommand{\arraystretch}{1.1}
\resizebox{\columnwidth}{!}{%
\begin{tabular}{cccc}
\toprule
\textbf{Algorithm} & \textbf{Mean} & \textbf{95\% Coverage Interval} & \textbf{Max $\eta$ (95\%)} \\
\midrule
Standard CG & 0.543 & [0.395, 0.715] & 14.04\% (observed) \\
Proposed PCG & 1.067 & [0.971, 1.187] & 1.02\% (observed)  \\
\bottomrule
\end{tabular}
\vspace{-0.8em}
}
\end{table}
Table \ref{tab:step_size} profiles the empirical first-step statistics. Standard CG exhibits a wide 95\% coverage interval [0.395, 0.715], and using its sample mean as a fixed step produces a maximum observed loss of 14.04\% within that central sample set. The proposed Jacobi-preconditioned direction has the narrower observed interval [0.971, 1.187], and the implemented $\alpha_{\rm fix}=1.075$ produces a maximum observed loss of 1.02\% in the same sense. Because the displayed interval endpoints are rounded, substituting them directly into \eqref{eq:sensitivity_law} gives the slightly more conservative value 1.15\%. These are sample statistics, not a deterministic bound; their role is to calibrate and test the hardware constant in the evaluated operating regime.

\subsection{Macroscopic SE and Power Trade-off}
Fig. \ref{fig:se_curves} illustrates the macroscopic user-average spectral efficiency (SE). At the high-SNR regime (32 dB), the proposed 6-bit Fixed-Step-Size BEM-PCG achieves 8.558 bps/Hz, closely approaching the two 12-bit high-resolution reference detectors and the 8.6 bps/Hz scheduled-modulation ceiling. Even at the medium-SNR regime (e.g., 26 dB), the maximum relative SE degradation compared to the 12-bit PIC-MMSE and MMSE references is 0.52\% and 0.31\%, respectively. Notably, the proposed scheme achieves a clear performance gain over the BEM-GS and dynamic BEM-CG baselines, which perform nearly identically. This suggests that our architecture largely recovers the 6-bit quantization loss while maintaining near-optimal macroscopic performance in the evaluated setting.

At low-to-medium SNRs, the fixed-step implementation slightly outperforms the dynamic-step BEM-PCG in the reported simulation. One possible explanation is that the calibrated $\alpha_{\text{fix}}$ acts as an implicit regularizer against instantaneous noise and quantization fluctuations. This is an empirical hypothesis rather than a consequence of the quadratic one-step theorem; controlled ablations would be required to establish the mechanism. The observation nevertheless shows that removing the $\mathcal{O}(U^2)$ step-size evaluator does not reduce SE in this evaluated regime.

\vspace{-1.2mm}

\subsection{Hardware Synthesis and Power Amortization}
To obtain an indicative hardware-efficiency estimate, the proposed equalizer and an MMSE reference are synthesized under SMIC 55nm CMOS at the typical-typical corner under 1.2\,V nominal supply using a multi-$V_t$ standard-cell library, both adopting 15-bit fixed-point quantization. Under block fading and low-to-medium mobility, the channel is assumed constant over the 14 OFDM symbols within one 5G NR slot \cite{3gpp2017ts38211}. For the considered per-subcarrier detector, the block-level factorization is reused across the slot, while only the symbol-level VSPU update is executed for each received symbol. Thus, the PCU and MIU costs are amortized over these 14 OFDM symbols, while the VSPU is accounted for directly.

\begin{table}[!t]
\caption{Amortized Power Breakdown and Indicative System-Level Savings}
\label{tab:power_synthesis}
\centering
\renewcommand{\arraystretch}{1.2}
\resizebox{\columnwidth}{!}{%
\begin{tabular}{ccc} 
\toprule
\textbf{Module} & \textbf{12-bit MMSE (mW)} & \textbf{6-bit PCG (mW)} \\
\midrule
PCU & 171.6 & 171.6 \\
MIU & 98.3 & 98.3 \\
VSPU & 1.0 & 164.3 \\
\midrule
Baseband Total & 270.9 & 434.2 \\
DSP Overhead & Base & +163.3 \\ 
\midrule
RF Front-End & Base & -452.4 \\ 
\midrule
Net Saving (proj.) & Base & -289.1 \\
\bottomrule
\end{tabular}
}
\vspace{-0.8em}
\end{table}

Table \ref{tab:power_synthesis} details the amortized breakdown. For the MMSE reference, the symbol-level post-processing is minimal and most computations are absorbed into the block-level PCU/MIU path, which is why the VSPU entry is negligible. By replacing the $\mathcal{O}(U^2)$ dynamic step-size evaluator with an $\mathcal{O}(U)$ element-wise preconditioning array, our VSPU consumes 164.3 mW. In addition, a model-based front-end power reduction of 452.4 mW is estimated for the 6-bit ADC setting \cite{medina2023vco, xu2024apccas}. Under these assumptions, our evaluation indicates a projected net system-level power reduction of 289.1 mW, highlighting the hardware potential of the proposed co-design.
\vspace{0.5mm}
\section{Conclusion}
In this paper, we alleviated the computation bottleneck of a BEM-inspired clipping-recovery surrogate for massive MIMO. We proved an exact finite-dimensional Jacobi normalization identity, the exact loss caused by a one-step fixed relaxation, and verifiable spectral conditions for uniform descent. These results support a calibrated one-step design that removes the $\mathcal{O}(U^2)$ step-size evaluator and enables an $\mathcal{O}(U)$ symbol-level feed-forward datapath, without claiming convergence of a multi-iteration PCG method or optimality for the exact quantized likelihood. In our evaluated setup, the proposed detector closely approaches the 12-bit reference baselines and 6-bit BEM-GS detection baseline. The co-design indicates a projected net power reduction of 289.1 mW under our evaluation assumptions.

\appendices

\section{Proofs of Propositions \ref{prop:spd_residual} and \ref{prop:line_search}}
\label{app:exact_descent}
For any nonzero $\mathbf{v}$,
\begin{equation}
\mathbf{v}^H\mathbf{A}\mathbf{v}
=\|\mathbf{H}\mathbf{v}\|_2^2+\mathbf{v}^H\mathbf{D}\mathbf{v}>0,
\end{equation}
because $\mathbf{D}=\mathbf{W}+\sigma^2\mathbf{I}\succ0$. Hence $\mathbf{A}\succ0$. Expanding $J$ and discarding terms independent of $\mathbf{x}$ gives
\begin{equation}
J(\mathbf{x})=\mathbf{x}^H\mathbf{A}\mathbf{x}
-2\operatorname{Re}(\mathbf{x}^H\mathbf{b})+\mathrm{const},
\end{equation}
so its unique minimizer is $\mathbf{A}^{-1}\mathbf{b}$. Moreover,
\begin{align}
\mathbf{b}-\mathbf{A}\hat{\mathbf{x}}^{(0)}
&=\mathbf{H}^H\mathbf{y}+\mathbf{W}\bar{\mathbf{x}}
-(\mathbf{G}+\mathbf{D})\hat{\mathbf{x}}^{(0)} \notag\\
&=\mathbf{W}\bar{\mathbf{x}}-\mathbf{D}\hat{\mathbf{x}}^{(0)},
\end{align}
where $\mathbf{G}\hat{\mathbf{x}}^{(0)}=\mathbf{H}^H\mathbf{y}$. This proves Proposition \ref{prop:spd_residual}.

For Proposition \ref{prop:line_search}, set
\begin{equation}
s=\mathbf{r}^H\mathbf{M}^{-1}\mathbf{r}>0,
\qquad
t=\mathbf{r}^H\mathbf{M}^{-1}\mathbf{A}\mathbf{M}^{-1}\mathbf{r}>0.
\end{equation}
Direct substitution yields, for real $\alpha$,
\begin{equation}
f(\alpha)=f(0)-2\alpha s+\alpha^2t.
\label{eq:app_quadratic}
\end{equation}
Thus $\alpha^*=s/t$. Completing the square gives
\begin{equation}
f(\alpha)-f(\alpha^*)=t(\alpha-\alpha^*)^2,
\qquad
f(0)-f(\alpha^*)=t(\alpha^*)^2,
\end{equation}
which proves \eqref{eq:sensitivity_law}. Finally,
$f(\alpha)-f(0)=\alpha t(\alpha-2\alpha^*)$, proving the necessary-and-sufficient descent condition \eqref{eq:residual_descent}. \hfill$\square$

\section{Proof of Proposition \ref{prop:unscaled_fragility}}
\label{app:unscaled_fragility}
For $\mathbf{A}=\mathbf{P}$, $\mathbf{r}=\mathbf{e}_i$, and the unpreconditioned direction $\mathbf{p}=\mathbf{r}$, direct substitution into the one-dimensional quadratic gives
\begin{equation}
f(\alpha)=f(0)-2\alpha+\alpha^2p_i.
\end{equation}
Hence $\alpha_i^*=1/p_i$ and, by the same completed-square calculation as in Appendix \ref{app:exact_descent}, $\eta_i=(\alpha p_i-1)^2$. For a fixed $\alpha$, the maximum over $p\in[p_{\min},p_{\max}]$ occurs at an endpoint. The minimax choice equalizes the two endpoint magnitudes,
\begin{equation}
1-\alpha p_{\min}=\alpha p_{\max}-1,
\end{equation}
which gives $\alpha=2/(p_{\min}+p_{\max})$ and \eqref{eq:unscaled_minimax}. For the strongest-user direction, \eqref{eq:residual_descent} reduces to $0<\alpha<2/p_{\max}$, proving the final statement. \hfill$\square$

\section{Proof of Theorem \ref{thm:jacobi_identity}}
\label{app:jacobi_identity}
Because $\mathbf{Z}$ has nonzero columns, $C_{ii}>0$, and
\begin{equation}
M_{ii}=A_{ii}=p_iC_{ii}+d_i>0.
\end{equation}
The diagonal entries of $\widetilde{\mathbf{A}}$ are therefore one. For $i\ne j$,
\begin{align}
[\widetilde{\mathbf{A}}]_{ij}
&=\frac{\sqrt{p_ip_j}\,C_{ij}}
{\sqrt{(p_iC_{ii}+d_i)(p_jC_{jj}+d_j)}} \notag\\
&=\sqrt{\gamma_i\gamma_j}\,T_{ij}.
\end{align}
Since $T_{ii}=1$, these entrywise identities are exactly equivalent to \eqref{eq:jacobi_identity}. Submultiplicativity of the spectral norm gives
\begin{align}
\|\widetilde{\mathbf{A}}-\mathbf{I}\|_2
&\le \|\boldsymbol{\Gamma}^{1/2}\|_2^2
\|\mathbf{T}-\mathbf{I}\|_2 \\
&\le\|\mathbf{T}-\mathbf{I}\|_2,
\end{align}
because $0<\gamma_i\le1$. If $\mathbf{D}=\mathbf{0}$, then $\boldsymbol{\Gamma}=\mathbf{I}$, proving the stated exact cancellation. \hfill$\square$

\section{Proofs of Corollaries \ref{cor:fixed_step} and \ref{cor:favorable_limit}}
Let $\mathbf{q}=\mathbf{M}^{-1/2}\mathbf{r}$. Equation \eqref{eq:exact_alpha} becomes
\begin{equation}
\alpha^*(\mathbf{r})=
\frac{\mathbf{q}^H\mathbf{q}}
{\mathbf{q}^H\widetilde{\mathbf{A}}\mathbf{q}}.
\label{eq:app_alpha_rayleigh}
\end{equation}
When $\varepsilon<1$, Theorem \ref{thm:jacobi_identity} places every eigenvalue of $\widetilde{\mathbf{A}}$ in $[1-\varepsilon,1+\varepsilon]$. The Rayleigh quotient in \eqref{eq:app_alpha_rayleigh} proves \eqref{eq:alpha_spectral_interval}. Writing
\begin{equation}
\lambda_{\mathbf{q}}=
\frac{\mathbf{q}^H\widetilde{\mathbf{A}}\mathbf{q}}
{\mathbf{q}^H\mathbf{q}}\in[1-\varepsilon,1+\varepsilon]
\end{equation}
gives $\eta=(\alpha_{\rm fix}\lambda_{\mathbf{q}}-1)^2$. This convex quadratic attains its maximum over the interval at an endpoint, proving \eqref{eq:fixed_step_bound}. Equalizing the endpoint errors yields $\alpha_{\rm fix}=1$ and worst-case loss $\varepsilon^2$. Since $1+\varepsilon<2$, the unit step satisfies the strict descent condition; the same argument gives the uniform condition \eqref{eq:uniform_descent}.

If $\|\mathbf{T}-\mathbf{I}\|_2\to0$, inequality \eqref{eq:jacobi_contraction} gives $\|\widetilde{\mathbf{A}}-\mathbf{I}\|_2\to0$. The interval \eqref{eq:alpha_spectral_interval} then contracts to one, uniformly over all nonzero $\mathbf{q}$ and hence all nonzero residuals. \hfill$\square$

\section{Proof of Proposition \ref{prop:minimax_interval}}
Let $\lambda=1/\alpha^*$. Then $\lambda\in[1/u,1/\ell]$ and \eqref{eq:sensitivity_law} becomes $\eta=(\alpha\lambda-1)^2$. The maximum absolute error over this interval occurs at an endpoint. The minimax positive step equalizes the endpoint magnitudes,
\begin{equation}
1-\frac{\alpha}{u}=\frac{\alpha}{\ell}-1,
\end{equation}
which gives $\alpha=2\ell u/(\ell+u)$ and endpoint error $(u-\ell)/(u+\ell)$. Squaring proves \eqref{eq:minimax_interval}. \hfill$\square$

\makeatletter
\let\originalthebibliography\thebibliography
\renewcommand{\thebibliography}[1]{%
  \originalthebibliography{#1}%
  \setlength{\itemsep}{2.5pt plus 0.2pt minus 0.1pt}}
\makeatother
\bibliographystyle{IEEEtran}
\bibliography{IEEEabrv, references}

\end{document}